\documentstyle[12pt]{article}
\newcommand{\beq}{\begin{equation}}
\newcommand{\eeq}{\end{equation}}
\newcommand{\beqa}{\begin{eqnarray}}
\newcommand{\eeqa}{\end{eqnarray}}
\newcommand{\bsquare}{\hbox{\rule{6pt}{6pt}}}

\begin{document}
\title{{\bf Non-perturbative Non-integrability}\\ 
    {\bf of Non-homogeneous Nonlinear Lattices}\\
       {\bf  Induced  by Non-resonance Hypothesis} 
\footnote{To appear in  {\it Physica}{\bf D}}}  
\author{{\bf Ken Umeno} 
\thanks{E-mail: chaosken@giraffe.riken.go.jp}\\
Laboratory for Information Representation\\
 Frontier Research Program, RIKEN\\ 
2-1 Hirosawa, Wako, Saitama 351-01, Japan} 
\date{}
\maketitle
\begin{abstract}
\normalsize
We have succeeded in  applying   Ziglin's test on non-integrability to  
non-homogeneous nonlinear lattices  
(Fermi-Pasta-Ulam lattices).
By explicit calculations of the eigenvalues of the 
monodromy matrices concerning the normal variational equations of 
Lam\'e type 
and with the use of 
 the non-resonance hypothesis about the eigenvalues, 
 we obtained a theorem proving  
 the nonexistence of additional analytic conserved quantities 
other than the Hamiltonian itself for 
  FPU lattices 
in the low energy limit. 
Furthermore, after introducing a concept of degree of 
non-integrability, 
we have investigated the classification of non-homogeneous 
nonlinear lattices using a transformation \(\Re\)   
from a non-homogeneous nonlinear lattice
  to another     
 non-homogeneous nonlinear lattice, which preserves their degree of 
 non-integrability. 
\end{abstract}

\vspace{0.5cm}
{\bf key words}: non-integrability, nonlinear lattices, singularity analysis,
  monodromy matrices, non-resonance condition 
\clearpage

\section{Introduction}
\setcounter{equation}{0}

Recently, we know two major sources of detecting non-integrability 
of Hamiltonian systems. One is Poincar\'e's theorem\cite{4poin} 
 of nineteenth century and the other is 
Ziglin's analysis\cite{zig} relatively recently found in the connection with 
 the classical singularity analysis\cite{rgb} by 
Fuchs, Kowalevskaya and Painlev\'e. What matters here is the following thing. 
The two different methods have  
significantly different  predictive capacities to tell non-integrability.  
While Poincar\'e's theorem is asserting the non-existence of additional 
conserved quantities such as 
\beq
\label{eq:pphi} 
  \Phi_{0}(\{q_{\nu}\})+\mu \Phi_{1}(\{q_{\nu},p_{\nu}\}) +
  \mu^{2} \Phi_{2}(\{q_{\nu},p_{\nu}\})+\cdots,
\eeq
 for a Hamiltonian system perturbed from an integrable systems
\beq
  H=H_{0}(\{q_{\nu}\})+\mu H_{1}(\{q_{\nu},p_{\nu}\}) +
  \mu^{2} H_{2}(\{q_{\nu},p_{\nu}\})+\cdots,
\eeq
 where \(\mu\) is a coupling constant whose explicit value is unknown,
 Ziglin's type theorem can prove the non-existence of additional 
analytic integral for an explicit Hamiltonian.
To illustrate the differences between them  more concretely, 
we can give the following examples.
The restricted three-body problem was proved to be non-integrable 
by Whittaker in 1904\cite{3whittaker} based on Poincar\'e's theorem, 
while Ziglin's theorem have never proved its non-integrability 
because of the problem that a 
Hamiltonian of the restricted three-body problem cannot be  defined 
without the infinitesimal operation \((m\rightarrow 0)\) and we don't 
have any 
explicit Hamiltonian which is necessary to perform  solid Ziglin's analysis. 
On the other hand, Poincar\'e's theorem assumes the condition on the 
Hessian of Hamiltonian, i.e.,
the Hessian of Hamiltonian 
should not vanish 
\beq
   |\frac{\partial^{2} H_{0}}{\partial q_{\nu} \partial q_{\lambda}}|
   \neq 0.
\eeq

By this assumption, Poincar\'e's theorem cannot not tell 
anything about the non-integrability of nonlinear lattices consisting of
harmonic  terms \(H_{0}=\frac{1}{2}\sum_{i=1}^{n}q^{2}_{i}\)
plus unharmonic terms such as the nonlinear 
lattices of Fermi-Pasta-Ulam type (FPU lattices). Therefore, in the present 
paper, we examine the possibility of non-integrability proof of 
FPU lattices via Ziglin's analysis. 
It should be also mentioned that Ziglin's analysis  
 have some technical difficulties, as is illustrated later, and  
no  non-integrability proof has been so far
studied either for realistic Hamiltonian systems
such as  the FPU lattices with non-homogeneous 
potential functions,
except few rare instances 
such as the Henon-Heiles systems with two degrees of 
freedom\cite{henon,zig1,ito} 
or for Hamiltonian systems with many degrees of freedom besides the case of  
 Hamiltonian systems with global
and symmetric coupling\cite{ku1}.

The purpose of the present paper is to show that 
 this  Ziglin's analysis 
can be performed for checking the non-integrability of  
 nonlinear lattices including the FPU lattices and to show that 
these nonlinear  lattices have no other 
analytic conserved 
quantities.

In Section 2,
we show that singularity analysis can be 
done for a class of FPU lattices by 
giving the explicit eigenvalues of the monodromy matrices about a 
special solution in terms of  elliptic functions.  
  
In Section 3,
we prove a  
non-integrability theorem for  FPU lattices 
based on the singularity analysis in Section 2,
which shows  that  
each FPU lattice has no additional analytic conserved quantities in the  
low energy limit.  
In Section 4 we discuss the classification of 
non-homogeneous nonlinear lattices via the degree of non-integrability. 
  
In Section 5, we give a brief summary and  open questions about the present 
analysis.

\section{Singularity analysis of nonlinear lattices}
\setcounter{equation}{0}

 Consider the following one-dimensional lattice:
\begin{equation}
\label{eq:general}
  H=\frac{1}{2}\sum_{i=1}^{n}p_{i}^{2}+
  \sum_{i=1}^{n+1}\upsilon(q_{i-1}-q_{i}).
\end{equation}
 If  this Hamiltonian admits the reflective invariance, namely, if it is 
 invariant under the involutive symplectic diffeomorphism,
\beq
  J_{r}H=H,
\eeq
 where
\begin{equation}
 J_{r}: \mbox{\boldmath$C$}^{2n}\rightarrow \mbox{\boldmath$C$}^{2n}, 
 J_{r}:(q_{1},\cdots,q_{n},p_{1},\cdots
 ,p_{n})\mapsto (-q_{1},\cdots,-q_{n},-p_{1},\cdots,-p_{n}),
\end{equation} 
 the potential function \(\upsilon(X)\) must satisfy the condition given by  
\beq
  \upsilon(X)=\upsilon(-X).
\eeq
 In the case of harmonic lattices, we have 
 \(\upsilon(X)=\frac{1}{2}\mu_{2}X^{2}\). 
 Thus more 
 general nonlinear lattices with the reflection symmetry can be defined 
 by the 
 following 
 polynomial function of even degree in \(X\):
\beq
\label{eq:potential}
  \upsilon(X)=\frac{\mu_{2}}{2}X^{2}+
  \frac{\mu_{4}}{4}X^{4}+\cdots +\frac{\mu_{2m}}{2m}
  X^{2m}.
\eeq
In this paper, however,  we assume that this potential function is a polynomial 
  function of degree  \(4\), namely, we consider the   
 Fermi-Pasta-Ulam(FPU) lattices\cite{fpu}
\begin{equation}
\label{eq:ulam}
  H_{FPU}=\frac{1}{2}\sum_{i=1}^{n}p_{i}^{2}+\frac{\mu_{2}}{2}
  \sum_{i=1}^{n+1}(q_{i-1}-q_{i})^{2}+
  \frac{\mu_{4}}{4}\sum_{i=1}^{n+1}(q_{i-1}-q_{i})^{4}.
\end{equation}

Imposing the fixed boundary condition as 
\beq
\label{eq:fixbound}
  q_{0}=q_{n+1}=0, \quad n=\mbox{odd},
\eeq
the equations of motion are given by
\begin{equation}
\label{eq:nls}
\begin{array}{l}
  \ddot{q_{1}}=  -\mu_{2}(2q_{1}-q_{2})-\mu_{4}(q_{1}^{3}+(q_{1}-q_{2})^{3}),\\
  \ddot{q_{2}}=  -\mu_{2}(2q_{2}-q_{1}-q_{3})-\mu_{4}((q_{2}-q_{1})^{3}+
   (q_{2}-q_{3})^{3}),\\
  \vdots  \cdots\\
  \vdots  \cdots\\
  \ddot{q_{n}}=  
  -\mu_{2}(2q_{n}-q_{n-1})-\mu_{4} ((q_{n}-q_{n-1})^{3}+q_{n}^{3})
\end{array}
\end{equation}

Define \(M'=\{(\mbox{\boldmath$q,p$})\in 
\mbox{\boldmath$C$}^{2n}|q_{1}\ne 0 \mbox{ or } p_{1}\ne 0\}\), 
\(\hat{M}=M'/J\) and 
the canonical projection \(\pi:M'\rightarrow \hat{M}\).  
An additional complex analytic integral in \(\mbox{\boldmath$C$}^{2n}\),
 if it exists, also induces the corresponding analytic integral  
 on \(\hat{M}\). Therefore it is sufficient to check the 
 non-integrability of  the reduced equations of  
 FPU lattices on \(\hat{M}\) in order to confirm the
 non-integrability of the original equations of FPU lattices.

Consider the following one-parameter family of the solution for (\ref{eq:nls}) 
\begin{equation}
\label{eq:solsol1}
\Gamma(\epsilon,t):  q_{j}=\frac{((-1)^{j}-1)}{2}(-1)^{\frac{j+1}{2}}C\phi(t)
\end{equation}
 with the initial condition 
\begin{equation}
\label{eq:initial}
    \phi (0)=1, \quad \dot{\phi}(0)=0,\quad H_{FPU}(0)=\epsilon.
\end{equation}
This is a complex solution in the complex time plane as   
\beq
\label{eq:compsol}
 q_{1}=C\phi(t),q_{2}=0,q_{3}=-C\phi(t),\cdots,q_{n-1}=0,
q_{n}=(-1)^{\frac{n-1}{2}}
 C\phi(t),
\eeq
 which is depicted in Fig. 1.
  Thus, the underlying equation of \(\phi (t)\) is reduced to the 
  following Hamiltonian 
  system with one degree of freedom:
\begin{equation}
\label{eq:oneh}
  \ddot{\phi}+2\mu_{2}\phi+2\mu_{4} C^{2}\phi^{3}=0,
\end{equation}
whose Hamiltonian is given by  
\begin{equation}
\label{eq:determin}
   H(\phi,\dot{\phi})=
  \frac{1}{2}(\dot{\phi})^{2}+\mu_{2}\phi^{2}+\frac{\mu_{4} C^{2}}{2}\phi^{4}
  =\mbox{Const.}
\end{equation}
 Then the original total energy \(\epsilon\) is written as follows: 
\begin{equation}
\label{eq:energyr}
  \epsilon=H=
  H(\phi,\dot{\phi})\frac{n+1}{2}C^{2}=
  \frac{n+1}{2}C^{2}(\mu_{2}+\frac{1}{2}\mu_{4} C^{2})
\end{equation} 
by  the initial condition (\ref{eq:initial}).
This equation (\ref{eq:energyr}) enable us to choose 
 \(C\) as the following {\it complex} quantity: 
\beq
\label{eq:cdterm} 
  C=\sqrt{\frac{\sqrt{\mu^{2}_{2}+\frac{4\epsilon}{n+1}\mu_{4}}
  -\mu_{2}}
  {\mu_{4}}}.
\eeq 
Therefore, \(H(\phi,\dot{\phi})\) becomes 
\beq
  H(\hat{q},\hat{p})=\frac{1}{2}\hat{p}^{2}+\mu_{2}\hat{q}^{2}
  +\frac{\sqrt{\mu_{2}^{2}+\frac{4\epsilon}{n+1}\mu_{4}}-\mu_{2}}{2}\hat{q}^{4},
\eeq
 where \(\hat{q}=\phi,\hat{p}=\dot{\phi}\).  
By combining (\ref{eq:determin}) with (\ref{eq:energyr}), we obtain 
the  
differential equation of \(\phi(t)\) as 
\begin{equation}
\label{eq:finalde}
 \frac{1}{2}(\dot{\phi})^{2}=\gamma_{2}
 (1-\phi^{2})+\frac{\gamma_{4}}{2}(1-\phi^{4})
\end{equation}
with the  system parameters
\begin{equation}
  \gamma_{2}\equiv \mu_{2},\quad \gamma_{4}\equiv \mu_{4}C^{2}.
\end{equation} 
Thus, the phase curve \(\Gamma(t,\epsilon)\)(\ref{eq:solsol1}) is given by 
the {\it elliptic} integral
\beq
\int_{\phi(0)=1}^{\phi(t)}
\frac{d\phi}{\sqrt{2}\sqrt{\gamma_{2}
(1-\phi^{2})+\frac{\gamma_{4}}{2}(1-\phi^{4})
}}=t
\eeq
 and 
the solution of  
this differential equation (\ref{eq:finalde}) with the condition 
\beq
 \gamma_{2m=4}\ne 0
\eeq
 is given explicitly by the formula  
\begin{equation}
  \phi(t)=cn(k;\alpha t),
\end{equation}
 where
\begin{equation}
\label{eq:3rel1}
   \alpha=\sqrt{2\gamma_{2}+2\gamma_{4}},\quad 
   k=\sqrt{\frac{\gamma_{4}}{2\gamma_{2}+2\gamma_{4}}},
\end{equation} 
\(cn(k;\alpha t)\) is the  Jacobi 
{\it cn} elliptic 
function, and \(k\) is  the modulus of the elliptic integral, because 
of 
the following equalities 
\beq
\begin{array}{l}
(\frac{d\phi}{dt})^{2}=\{-\alpha sn(k;\alpha t)dn(k;\alpha t)\}^{2}\\
		      =\alpha^{2}\{(-2k^{2}+1)(1-cn^{2}(k;\alpha t))
		      +k^{2}(1-cn^{4}(k;\alpha t))\}\\
		      =2\gamma_{2}(1-\phi^{2})+\gamma_{4}(1-\phi^{4})
\end{array}
\eeq
in Eq. (\ref{eq:finalde}).
We remark here that 
the relation
\beq
 \gamma_{2}+\gamma_{4}=\mu_{2}+C^{2}\mu_{4}=
 \sqrt{\mu_{2}^{2}+\frac{4\epsilon}{n+1}\mu_{4}}>0,
\eeq
holds 
for \(\mu_{4}>0,\mu_{2}> 0,\epsilon>0\). This means that  
the modulus of the elliptic function \(k\) satisfies the following relation:
\beq
\label{eq:modulus} 
  0< k=\frac{1}{\sqrt{2}}
\sqrt{1-\frac{1}{\sqrt{1+\frac{4\epsilon\mu_{4}}{(n+1)\mu^{2}_{2}}}}} 
<\frac{1}{\sqrt{2}} 
\eeq
 and \(C\) is a real quantity. 
On the contrary, if the following relations 
\beq
 \mu_{2}>0, \mu_{4}>0, \epsilon<0 
\eeq
are satisfied,   
the modulus of the elliptic function \(k\) becomes pure imaginary, i.e., 
  \(k^{2}<0\).  \(C\) becomes also pure imaginary as in Fig.1. We note here
that even in this {\it unphysical} case, there is no barrier to perform 
the following singularity analysis. However, in the following we shall
consider the physical case with the condition as 
\(\mu_{2}>0,\mu_{4}>0, \epsilon>0\).    
According to the reduction \(\pi : M'\rightarrow \hat{M}=M'/J\) by the 
reflective symmetry \(J\),   
 the solution (\ref{eq:compsol})
of the induced FPU lattices on \(\hat{M}\) for \(\mu_{4}>0,\mu_{2}\geq 0\)
is a single-valued, meromorphic and doubly periodic function in \(t\), 
having the two {\it fundamental periods} in the complex 
time plane as follows: 
\begin{equation}
\label{eq:rel2}
    T_{1}(\epsilon,\mu)=\frac{2K(k)}{\alpha},
    \quad T_{2}(\epsilon,\mu)=\frac{2K(k)+2iK'(k)}{\alpha}.
\end{equation}    
 Here \(K(k)\) and \(K'(k)\) are 
 the complete elliptic integrals of the first kind: 
\begin{equation}
   K(k)=\int_{0}^{1}\frac{dv}{\sqrt{(1-v^{2})(1-k^{2}v^{2})}},\quad
   K'(k)=\int_{0}^{1}\frac{dv}{\sqrt{(1-v^{2})(1-(1-k^{2})v^{2})}}.
\end{equation} 
 Note here that these solutions have a {\it single} pole at 
 \(t=\tau,\) where 
 \(\tau=\frac{2K(k)}{\alpha}+i\frac{K'(k)}{\alpha}\quad (\mbox{mod }
 T_{1},T_{2})\)
 in the parallelogram of each period cell; the phase curves 
 \(\Gamma(\epsilon,t)\) are punctured tori. 
If we consider the solution (\ref{eq:compsol}) not on the 
reduced phase space \(\hat{M}=M'/J\) but on the original 
phase space \(M'\), the solution  has the following fundamental periods as
\beq
 T'_{1}(\epsilon,\mu)=\frac{4K(k)}{\alpha},
    \quad T'_{2}(\epsilon,\mu)=\frac{2K(k)+2iK'(k)}{\alpha}, 
\eeq
 which are different from (\ref{eq:rel2}) and this solution on 
\(M'\) has 
{\it two}  poles at \(t=\tau',\)\\ where 
 \(\tau'=\frac{2K(k)}{\alpha}+i\frac{K'(k)}{\alpha},
\frac{4K(k)}{\alpha}+i\frac{K'(k)}{\alpha}
\quad (\mbox{mod }
 T'_{1},T'_{2})\)
in the parallelogram of each periodic cell.

 Let 
 us consider the variational equations along the phase curve 
 \(\Gamma(\epsilon,t)\) in (\ref{eq:solsol1}) of the general 
 nonlinear lattices with the reflection symmetry.  
 The linearized variational equations are written as follows:
\begin{equation}
\label{eq:firstve}
\begin{array}{l}
  \dot{\eta_{j}}= 
  \ddot{\xi_{j}}=
  -\sum_{k=1}^{n}\left.\frac{\partial^{2}V}{\partial q_{k}\partial q_{j}}
  \right|_{\Gamma}
  \xi_{k}\\
  =
 -(\gamma_{2}+3\gamma_{4}cn^{2}(k;\alpha t))
 (2\xi_{j}-\xi_{j-1}-\xi_{j+1})
  \mbox{  for  }1\leq j\leq n,
\end{array}
\end{equation}
 where
   \(\xi_{0}=\xi_{n+1}=\eta_{0}=\eta_{n+1}=0\) and \(\xi_{j}=\delta q_{j},
   \eta_{j}=\delta p_{j}\quad(1\leq j\leq n)\).\\
  If we rewrite  
  these linear 
  variational equations (\ref{eq:firstve}) in the  form of the vector 
   variational equation as
\begin{equation}
\frac{d^{2}}{dt^{2}}
\mbox{\boldmath$\xi$}=-(\gamma_{2}+3\gamma_{4}cn^{2}(k;\alpha t))
\left[
\begin{array}{ccccc}
2 & -1 & 0 & \cdots & 0 \\
-1 & 2 & -1 & \cdots & 0 \\
0 & -1 & 2 & -1 & \cdots \\
\cdots & \cdots & \cdots & \cdots & \cdots \\
0 & \cdots & 0 & -1 & 2,  
\end{array}
\right]
\mbox{\boldmath$\xi$}
\end{equation}
and use the fact that  the eigenvalues of the \(n\times n\) symmetric matrix 
\begin{equation}
\mbox{\boldmath$G$}=
\left[
\begin{array}{ccccc}
2 & -1 & 0 & \cdots & 0 \\
-1 & 2 & -1 & \cdots & 0 \\
0 & -1 & 2 & -1 & \cdots \\
\cdots & \cdots & \cdots & \cdots & \cdots \\
0 & \cdots & 0 & -1 & 2  
\end{array}
\right]
\end{equation}
are obtained as \(\{4\mbox{sin}^{2}(\frac{j\pi}{2(n+1)})|1\leq 
  j\leq n\}\) 
 by a normal orthogonal transformation 
\(\mbox{\boldmath$G$}\rightarrow
\mbox{\boldmath$OGO^{-1}$}\),  
 the variational equations (\ref{eq:firstve}) can be transformed  
 into the decoupled form:
\begin{equation}
\label{eq:cve}
  \ddot{\xi'_{j}}(t)=-4\mbox{sin}^{2}(\frac{j\pi}{2(n+1)})
  (\gamma_{2}+3\gamma_{4}cn^{2}(k;\alpha t))
  \xi'_{j}(t)\quad(1\leq j \leq n),
\end{equation}
where \(\mbox{\boldmath$\xi'$}=\mbox{\boldmath$O\xi$}\).
Clearly, these equations are 
regarded as  the vector form of {\it Hill's equations}\cite{3hill}
\beq
  \frac{d^{2}\mbox{\boldmath$\xi'$}}{dt^{2}}+\mbox{\boldmath$A$}(t)
  \mbox{\boldmath$\xi'$}=0,\quad
  \mbox{\boldmath$A$}(t+T)=\mbox{\boldmath$A$}(t),
\eeq
 where \(T=T_{1},T_{2}\).  
For \(j=\frac{n+1}{2}\), the relation  
\begin{equation}
\label{eq:mannaka}
\xi'_{\frac{n+1}{2}}=\sqrt{\frac{2}{n+1}}(\xi_{1}-\xi_{3}+\xi_{5}+\cdots+(-1)^{\frac{n-1}{2}}
\xi_{n})
\end{equation}
holds. The corresponding variational equation 
\begin{equation}
\label{eq:tangenve}
  \ddot{\xi'}_{\frac{n+1}{2}}=-2(\gamma_{2}+3\gamma_{4}cn^{2}(k;\alpha t))
  \xi'_{\frac{n+1}{2}}(t)
\end{equation} 
has a time-dependent integral 
\(I(\mbox{\boldmath$\xi$},\dot{\mbox{\boldmath$\xi$}};t)
\equiv I(\mbox{\boldmath$\xi$},\mbox{\boldmath$\eta$};t)\) 
derived  from the variational 
operation  \(\mbox{\boldmath$D$}\) applied to the Hamiltonian
(\ref{eq:ulam}):
\beq
\begin{array}{l}
  I(\mbox{\boldmath$\xi$},\mbox{\boldmath$\eta$};t)=
   \mbox{\boldmath$D$}H\equiv 
  (\mbox{\boldmath$\eta$}\cdot
  \frac{\partial}{\partial\mbox{\boldmath$p$}}+\mbox{\boldmath$\xi$}\cdot
  \frac{\partial}{\partial\mbox{\boldmath$q$}})H
  =\mbox{\boldmath$\eta$}\cdot\mbox{\boldmath$p$}+\mbox{\boldmath$\xi$}
  \cdot\mbox{\boldmath$V_{q}$}\\
  =C\dot{\phi}(\eta_{1}-\eta_{3}+\eta_{5}+
  \cdots+(-1)^{\frac{n-1}{2}}\eta_{n})\\
  +2(C\gamma_{2}cn(k;\alpha t)+
  C\gamma_{4}cn^{3}(k;\alpha t))
  (\xi_{1}-\xi_{3}+\xi_{5}+\cdots+(-1)^{\frac{n-1}{2}}
  \xi_{n}),
\end{array}
\eeq
because
\beq
\begin{array}{l}
\frac{1}{C}\frac{dI}{dt}=\dot{\phi}(\ddot{\xi_{1}}-\ddot{\xi_{3}}+
\cdots+(-1)^{\frac{n-1}{2}}
\ddot{\xi_{n}})\\
+2\dot{\phi}(\gamma_{2}+3\gamma_{4}cn^{2}(k;\alpha t))
(\xi_{1}-\xi_{3}+\cdots+(-1)^{\frac{n-1}{2}}
\xi_{n})
=0. 
\end{array}
\eeq
Eq. (\ref{eq:tangenve}) is called the {\it  tangential variational 
equation}.
On the other hands, a \((2n-2)\)-dimensional 
{\it normal variational equation}(NVE) is defined by 
the equations  (\ref{eq:cve}) 
with the tangential variational equation (\ref{eq:tangenve}) 
removed as follows:
\begin{equation}
\label{eq:nveev}
\begin{array}{l}
  \dot{\eta'}_{j}=
  -4\mbox{sin}^{2}(\frac{j\pi}{2(n+1)})(\gamma_{2}+3\gamma_{4} 
  cn^{2}(k;\alpha t))
  \xi'_{j},\\
  \dot{\xi'}_{j}=\eta'_{j}
\end{array}
\eeq
for \(1\leq j(\ne \frac{n+1}{2})
  \leq n\).

 Let us  consider the  
monodromy matrices \(g\)   
 defined by the analytic continuation of the solution 
\(\zeta'(t)=(\xi'_{1}(t),\eta'_{1}(t),\cdots,\hat{\xi}'_{\frac{n+1}{2}}(t),
\hat{\eta}'_{\frac{n+1}{2}}(t),\cdots,\xi'_{n}(t),\eta'_{n}(t))\)
of the NVE (\ref{eq:nveev}) 
along the  periodic orbits in the phase curves \(\Gamma(\epsilon,t)\)
 as follows:
\begin{equation}
\label{eq:monodef4}
   \mbox{\boldmath$\zeta'$}(T_{1})=g_{1}\mbox{\boldmath$\zeta'$}(0),\quad 
   \mbox{\boldmath$\zeta'$}(T_{2})=g_{2}\mbox{\boldmath$\zeta'$}(0).
\end{equation}
 The periods in Eqs. (\ref{eq:monodef4}) are \(T_{1},T_{2}\) in 
Eqs. (\ref{eq:rel2}), 
 respectively.
 These two fundamental periods \(T_{1}\) and \(T_{2}\) naturally 
 form the parallelogram, whose associate monodromy matrices are given by 
 \(g_{1}g_{2}g^{-1}_{1}g^{-1}_{2}(\equiv g_{*})\). 
 These monodromy matrices are naturally endowed
 with the symplectic structure and the 
 pairing properties of the eigenvalues, namely 
 \(\{\sigma_{1},\sigma^{-1}_{1},\sigma_{2},\sigma^{-1}_{2},
    \cdots,\sigma_{n},\sigma^{-1}_{n}\}\).
 In general, the explicit calculation of the eigenvalues of 
 the monodromy matrices is an unsuccessful business 
 except rare cases such as Hamiltonian systems with 
 homogeneous polynomial functions\cite{yo}, {\it Riemann's equation}, and the 
 {\it Jordan-Pochhammer equations}\cite{3takano,3hejhal,3churchill}. 
 This is one of the 
 unavoidable difficulties
 in performing Ziglin's analysis of general dynamical systems. 
 In our case, however, we can compute explicitly    
 the eigenvalues of  
 the {\it commutator} \(g_{*}=g_{1}g_{2}g^{-1}_{1}g^{-1}_{2}\) 
 by using the fact that  
the normal variational equation (\ref{eq:nveev}) 
 happens  to be in a class of    
 {\it Lam\'e 
 equations} \cite{whit}
\begin{equation}
 \frac{d^{2} y}{dt^{2}}-(E_{1}sn^{2}(k;\alpha t)+E_{2})y=0,
\end{equation}
 where \(E_{1}\) and \(E_{2}\) are constants, and the eigenvalues \(\sigma\) 
 of the commutator \(g_{*}=g_{1}g_{2}g^{-1}_{1}g^{-1}_{2}\) are known
 \cite{whit} to be 
  determined by the 
 indicial equation 
\begin{equation}
\label{eq:indicial1}
    \Delta^{2}-\Delta-(\alpha^{2}k^{2})E_{1}=0,
    \quad \sigma=\mbox{exp}(2\pi i\Delta)
\end{equation}
 with the singular point (pole) \(\tau\) located at the center of 
  the parallelogram:  
\begin{equation}
\tau=\frac{T_{1}+T_{2}}{2}=\frac{2K(k)+iK'(k)}{\alpha}.
\end{equation}
 If  we apply  
 the indicial equation(\ref{eq:indicial1}) to the 
normal variational equations (\ref{eq:nveev}) 
of the FPU lattices, the exponents of 
the eigenvalues of \(g_{*}\)
are given by 
\begin{equation}
  \Delta^{2}-\Delta-12\frac{\gamma_{4}}{\alpha^{2}k^{2}}
  \mbox{sin}^{2}(\frac{j\pi}{2(n+1)})=0.
\end{equation} 
Noting   
\begin{equation}
  \frac{\gamma_{4}}{\alpha^{2}k^{2}}=1
\end{equation}
from (\ref{eq:3rel1}), we finally obtain the eigenvalues of the commutator as 
\begin{eqnarray}
\label{eq:fpumonodromy}
  \sigma_{j}=\mbox{exp}(2\pi i \frac{1\pm\sqrt{1+
  48\mbox{sin}^{2}\frac{j\pi}{2(n+1)}}}{2})\nonumber\\
	    =-\mbox{exp}(\pm \pi i\sqrt{25-24\mbox{cos}(\frac{j\pi}{n+1})})
\end{eqnarray}
for \(1\leq j(\ne \frac{n+1}{2})\leq n\), because there is {\it only one}
pole singularity inside the parallelogram formed by counterclockwise 
closed loop of the monodromy group \(g_{1}g_{2}g^{-1}_{1}g^{-1}_{2}\).
 Even for more general nonlinear 
lattices  given by  
 potential functions (\ref{eq:potential}) of degree \(2m\) 
 we can also compute the phase factors. 
Let us consider singular solutions  
  \(\phi(t)\) and \(\xi_{j}(t)\) whose singularities are located  
 at \(t=\tau\) in the complex time plane as follows: 
 \beq
 \label{eq:assumpsyg}
 \begin{array}{l}
   \phi(t)=C'(t-\tau)^{\beta},\quad \beta<0\nonumber\\
   \xi_{j}(t)=C''(t-\tau)^{\nu_{j}},
 \end{array}
 \eeq
 where \(C'\) is a particular constant, \(C''\) is an arbitrary constant. 
 Thus, we can easily obtaine the formulas  the relations  
 for \(\beta,C',\nu_{j}\) as  
\beq
\begin{array}{l}
  \beta=\frac{-1}{m-1},\\
  (C')^{2m-2}=-\frac{\beta(\beta-1)}{2\gamma_{2m}},\\
  \nu_{j}^{2}-\nu_{j}-2\frac{m(2m-1)}{(m-1)^{2}}\mbox{sin}^{2}(\frac{j\pi}
  {2(n+1)})=0
\end{array}
\eeq
by the equations of motion. 
 With the use of the fact that  the phase factor 
 is obtained by means of the analytic continuation along the 
 closed loop around the singular point \(t^{\nu_{j}}\), we obtain 
 the phase factors as  
\beq
\label{eq:generalmonodromy}
\begin{array}{c}
 \mbox{exp}(2\pi i \nu_{j})\\
 =\mbox{exp}\{2\pi i \frac{1\pm \sqrt{1+8\frac{m(2m-1)}{(m-1)^{2}}
 \mbox{sin}^{2}(\frac{j\pi}{2(n+1)}}}{2})\}\\
 =-\mbox{exp}\{\pm \pi i \sqrt{1+8\frac{m(2m-1)}{(m-1)^{2}}
 \mbox{sin}^{2}(\frac{j\pi}{2(n+1)})}\}
\end{array}
\eeq
 where \(2m\) is the degree of the potential polynomial  of \(
 \upsilon(X)\) in (\ref{eq:potential}). It is easy to confirm that in  
 the case that \(2m=4\), 
the  formula (\ref{eq:generalmonodromy}) recover the eigenvalues 
(\ref{eq:fpumonodromy}) of the monodromy matrices 
obtained from the {\it Jacobi elliptic function}. However, in the other 
case that \(2m>4\),  
we cannot compute the eigenvalues of monodromy matrices only from the 
phase factors in general, 
because the normal variational equations do not belong 
to a class of {\it Lam\'e equations}.
To summarize, we can say  
that  
it is essential 
to perform the singularity analysis towards non-homogeneous 
 nonlinear lattices using the explicit eigenvalues 
of the monodromy matrices that 
the special solutions can  be written in terms of elliptic functions and 
the normal variational equations are in a type of {\it Lam\'e equations}. 
 
\section{Non-integrability theorem}
\subsection{Non-integrability proof in the integrable limit} 
\setcounter{equation}{0}
The chief purpose of the present section is to give a theorem 
telling the non-integrability by using the monodromy matrices obtained 
in Section 2. We consider  monodromy matrices \(g\). 
If the eigenvalues \(\{\sigma_{1},\sigma^{-1}_{1},\cdots,
 \sigma_{n},\sigma^{-1}_{n}\}\)  
of monodromy matrices \(g\) do not satisfy the following relation 
\beq
  \sigma_{1}^{l_{1}}\sigma_{2}^{l_{2}}\cdots \sigma_{n}^{l_{n}}=1
\eeq
 for any set of integers \(\{l_{1},\cdots,l_{n}\}\) except the trivial 
case \(l_{1}=l_{2}=\cdots =l_{n}=0\), we call the monodromy matrices \(g\)  
{\it non-resonant}. 
It is already known that  
the existence of a non-resonant monodromy matrix is a basic 
assumption in order to perform Ziglin's analysis\cite{zig}.
Moreover, if there are {\it straight-line solutions} 
such as (\ref{eq:compsol}) whose monodromy matrices 
are non-resonant and if the variational equations can be diagonalized into 
a decoupled form like (\ref{eq:cve}),  
  Ziglin's theorem can be generalized to   
Yoshida's theorem for Hamiltonian systems composed by 
kinetic energy terms and potential energy terms\cite{yopd}.
 
Yoshida's theorem 
asserts the following in terms of two different monodromy matrices
\(\{g_{a},g_{b}\}\):
Suppose that there exists an additional complex analytic integral, which 
is holomorphic along the solution (\ref{eq:solsol1}), and that
one of the monodromy matrices 
\(g_{a}\) is non-resonant. Then it is necessary that   
 one of the following  two cases, namely, 
\begin{itemize} 
 \setlength{\itemsep}{0pt}
 \item[(I)] 
\(g_{b}(\lambda_{j})\)  must preserve the eigendirection of \(g_{a}(\lambda_{j})\)
, i.e., \(g_{a}(\lambda_{j})\) must commute with \(g_{b}(\lambda_{j})\), 
\item[(II)]
\(g_{b}(\lambda_{j})\)
 must permute the eigendirection of \(g_{a}(\lambda_{j})\), 
 i.e., \(g_{b}(\lambda_{j})\) is written by 
 \(\left[
 \begin{array}{cc}
   0 & \beta \\
   -\frac{1}{\beta} & 0 
 \end{array}
 \right]\) in the base of \(g_{a}\)  
 having the  
 eigenvalues \(i\) and \(-i\) 
\end{itemize}
 for some suffix \(j\), 
at least occurs 
for any other  monodromy matrix 
\(g_{b}\) represented in the 
basis of  
 \(g_{a}\). 
As for the non-resonance condition of the monodromy 
matrix \(g_{*}=g_{1}g_{2}g^{-1}_{1}g^{-1}_{2}\), the following 
lemma is already known:
\newtheorem{th2}{Lemma} 
\begin{th2}[\cite{4um0,4umth},1994]
\label{th3:fputheorem1} 
The n quantities
\(\{\sqrt{25-24\mbox{cos}\frac{j\pi}{n+1}}| j=1,\cdots,n\}\)
are rationally independent.
\end{th2}

We remark that 
   the rational independency of the set\\
\(\{\sqrt{25-24\mbox{cos}\frac{j\pi}{n+1}}|j=1,\cdots n\}\)
guarantees that
 the commutator \(g_{*}=g_{1}g_{2}g^{-1}_{1}g^{-1}_{2}\) whose 
eigenvalues are given in Eq. (\ref{eq:fpumonodromy}) is non-resonant.

For connection with the  algebraic number theory on the {\it cyclotomic field}
\(\mbox{\boldmath$Q$}(\mbox{exp}(\frac{\pi i}{n+1}))\) over
\(\mbox{\boldmath$Q$}\) see Ref. \cite{4umth}.
Thus, we can regard that the non-resonance hypothesis holds for 
the monodromy matrix  \(g_{*}\).
Now, using Yoshida's theorem and the variational 
analysis in the former section, we obtain a theorem which proves  
the non-integrability of a FPU lattice for an arbitrary set of 
the system parameters 
\(\{n,\mu_{2},\mu_{4}|n(\geq3):\mbox{odd},\mu_{2}>0,\mu_{4}>0\}\) as follows:
\newtheorem{th3}{Theorem}

\begin{th3}
\label{th3:fputheorem}
A FPU lattice (\ref{eq:ulam}) which is characterized by 
 an arbitrary set of the system parameters
\(\{n,\mu_{2},\mu_{4}|n \geq3, n:\mbox{odd},\mu_{2}>0,\mu_{4}>0\}\)
 has no  analytic first 
integrals besides the Hamiltonian itself   
for fully small energy \(\epsilon(\approx 0)\).
\end{th3}

({\bf Proof of Theorem \ref{th3:fputheorem1}})

  From the non-resonance hypothesis on \(g_{*}\), 
  we can apply   the Yoshida theorem\cite{yopd} to 
  the FPU lattices.
   According to the above argument, if we prove that at least one 
 monodromy matrix \(g_{s}\in\{g_{1},g_{2}\}\) 
does not have the following properties
 
\begin{itemize} 
 \setlength{\itemsep}{0pt}
\item[(a)] \(g_{s}(\lambda_{j})\) 
  preserves the eigendirection 
 of \(g_{*}(\lambda_{j})\) 
\\and   
\item[(b)] \(g_{s}(\lambda_{j})\) 
 permutes the eigendirection of \(g_{*}(\lambda_{j})\)(the eigenvalues of 
 \(g_{s}(\lambda_{j})\) are \(i,-i\)),
\end{itemize}
 at once  
 for any suffix \(j\)(\(1\leq j(\ne n+1/2) \leq n\)) of \(g_{s}(\lambda_{j})\), 
 then the 
 FPU lattices
 \cite{fpu} are concluded to  have 
 no other analytic conserved quantities besides the Hamiltonian 
 itself, i.e., the assertion of the present theorem holds. 
In the following, we will show that \(g_{1}(\lambda_{j})\) for any \(j(\ne 
\frac{n+1}{2})\) has neither the property (a) and the property (b).

When we take the 
 limit 
 \(\epsilon\rightarrow 0\), the relations 
\beq
\gamma_{4}=\mu_{4} C^{2},\cdots,\gamma_{2m}=
  \mu_{2m}C^{2m-2}\rightarrow 0,\quad \alpha
 \rightarrow \sqrt{2\mu_{2}},\quad  
 k\rightarrow 0
\eeq 
hold
  and we can compute the periods of the monodromy matrices 
 \(g_{1},g_{2}\) in the limit as 
\beq
 T_{1}\rightarrow \frac{1}{\sqrt{2\gamma_{2}}}\pi,\quad
 T_{2}\rightarrow \frac{1}{\sqrt{2\gamma_{2}}}\pi+i\infty
\eeq 
 by  the formula (\ref{eq:rel2}).  
To compute  the  monodromy matrix \(g_{1}(\lambda_{j})\) in this 
limit, we rewrite  
the variational equations (\ref{eq:nveev}) in terms of 
the modulus of the elliptic integral \(k\) as follows
\beq
\label{eq:simvar}
  \ddot{\xi'}_{j}+
  4\mu_{2}\mbox{sin}^{2}(\frac{j\pi}{2(n+1)})(1+\frac{2k^{2}}{1-2k^{2}}
   cn^{2}(k;\alpha t))
  \xi'_{j}=0,\quad j=1,\cdots,n
\eeq
by virtue of equations (\ref{eq:cdterm}) and (\ref{eq:3rel1}).
For arbitrary \(k\in\left[0,1\right]\),
we have a fundamental system of solutions as 
\(\{\xi^{a}_{j}(k;t),\xi^{b}_{j}(k;t)\}\) of (\ref{eq:simvar}) satisfying 
\beq
\label{eq:initcs}
\begin{array}{l}
\xi^{a}_{j}(k;0)=1,\quad \frac{d}{dt}\xi^{a}_{j}(k;t)|_{t=0}=
\dot\xi^{a}_{j}(k;0)=0\\
\xi^{b}_{j}(k;0)=0,\quad \frac{d}{dt}\xi^{b}_{j}(k;t)|_{t=0}=
\dot\xi^{b}_{j}(k;0)=1.
\end{array}    
\eeq

Because \(cn^{2}(k;\alpha t)\) in Eqs. (\ref{eq:simvar}) 
has the following Taylor expansions \cite{byrd} at 
\(\kappa \equiv k^{2}=0\): 
\beq
  cn^{2}(k;\alpha t) = \mbox{cos}^{2}(\frac{\alpha\pi}{2K} t) 
   -\frac{1}{2}\kappa \mbox{sin}^{2}(\frac{\alpha\pi}{2K} t)
    \mbox{cos}^{2}(\frac{\alpha\pi}{2K} t)+\cdots
\eeq 
and analytic in \(\kappa\) at \(\kappa=0\), the 
fundamental system of solutions \(\{\xi^{a},\xi^{b}\}\) are also analytic
 
 in \(\kappa\) at \(\kappa=0\) \cite{ito}  as 
\beq
\begin{array}{l}
  \xi^{a}_{j}(k;t)=\xi^{a}_{j,0}(t)+\xi^{a}_{j,1}(t)\kappa 
  +\xi^{a}_{j,2}(t)\kappa^{2}+\cdots\\
  \xi^{b}_{j}(k;t)=\xi^{b}_{j,0}(t)+\xi^{b}_{j,1}(t)\kappa 
+\xi^{a}_{j,2}(t)\kappa^{2}+\cdots,
\end{array}
\eeq
 where the unperturbed parts 
\(\{\xi^{a}_{j,0},\xi^{b}_{j,0}\}\) are obtained by 
\beq
\label{eq:fusy}
\begin{array}{l}
 \xi^{a}_{j,0}(t)=\mbox{cos}
 (2\mbox{sin}(\frac{j\pi}{2(n+1)})\sqrt{\mu_{2}}t),\\
  \xi^{b}_{j,0}(t)=\frac{1}{2\sqrt{\mu_{2}}\mbox{sin}(\frac{j\pi}{2(n+1)})}
\mbox{sin}(2\mbox{sin}(\frac{j\pi}{2(n+1)})\sqrt{\mu_{2}}t)
\end{array} 
\eeq
 from  the initial conditions (\ref{eq:initcs}). 
The monodromy matrix \(g_{1}(\lambda_{j})\) is given by 
\beq
  g_{1}(\lambda_{j})=\left[
  \begin{array}{cc}
  \xi^{a}_{j}(k;T_{1}) & \xi^{b}_{j}(k;T_{1})
 \\
\dot\xi^{a}_{j}(k;T_{1}) & \dot\xi^{b}_{j}(k;T_{1})
  \end{array}\right],
\eeq
 and 
with the use of the fact that 
the fundamental system of solutions (\ref{eq:fusy}) 
are analytic in \(\kappa\) at \(\kappa=0\), we show that 
 \(g_{1}(\lambda_{j})\) for \(j=1,\cdots, n\) have 
 also the Taylor expansions as
\beq 
g_{1}(\lambda_{j})=g_{1,0}(\lambda_{j})+g_{1,1}(\lambda_{j})\kappa
  +g_{1,2}(\lambda_{j})\kappa^{2}+\cdots.
\eeq
Thus now, when we consider 
the low energy limit (\(\epsilon\rightarrow 0\), \(\kappa\rightarrow 0\)), 
 we  obtain the formula 
\begin{equation}
  g_{1}(\lambda_{j})\rightarrow\left[
  \begin{array}{cc}
  \mbox{cos}(\sqrt{2}\pi\mbox{sin}(\frac{j\pi}{2(n+1)})) & 
\frac{1}{2\mbox{sin}(\frac{j\pi}{2(n+1)})\sqrt{\mu_{2}}}
 \mbox{sin}(\sqrt{2}\pi\mbox{sin}(\frac{j\pi}{2(n+1)}))
 \\
 -2\mbox{sin}(\frac{j\pi}{2(n+1)})\sqrt{\mu_{2}}
 \mbox{sin}(\sqrt{2}\pi\mbox{sin}(\frac{j\pi}{2(n+1)}))  
 & 
\mbox{cos}(\sqrt{2}\pi\mbox{sin}(\frac{j\pi}{2(n+1)})\pi)
  \end{array}\right]. 
\end{equation}
This means that   
the eigenvalues of \(g_{1}(\lambda_{j})\) tend to
\beq
\label{eq:limiteigenvalues} 
\{\mbox{exp}(i\pi(\sqrt{2}\mbox{sin}(\frac{j\pi}{2(n+1)})),
\mbox{exp}(-i\pi(\sqrt{2}\mbox{sin}(\frac{j\pi}{2(n+1)}))
\}
\eeq 
 and  
 \(g_{1}(\lambda_{j})\) for any \(j(\ne \frac{n+1}{2})\) 
 does not have the property of (b).\\

Now assume that 
 \(g_{2}(\lambda_{j})\) for 
  some 
 \(j(\ne\frac{n+1}{2})\) has 
 the property of (a). Then if \(g_{1}(\lambda_{j})\) 
 as well as \(g_{2}(\lambda_{j})\) has also the  
 property of (a), we have 
\beq
\label{eq:commono}
 g_{*}(\lambda_{j})=g_{1}(\lambda_{j})g_{2}(\lambda_{j})
 g^{-1}_{1}(\lambda_{j})g^{-1}_{2}(\lambda_{j})=
 \mbox{\boldmath$id$},
\eeq
 where \(\mbox{\boldmath$id$}\) denotes 
 the \(2\times 2\) identity matrix. This relation (\ref{eq:commono}) 
  means that \(g_{1}(\lambda_{j})\) 
 and \(g_{2}(\lambda_{j})\) commute each other and clear contradicts  
 the non-resonance hypothesis of \(g_{*}\). 
Consider the other case where  
  \(g_{1}(\lambda_{j})\) has the property of (a)   
  and \(g_{2}(\lambda_{j})\) does not have the property of (a).
  However, in the representation of \(g_{1}(\lambda_{j})\) and 
 \(g_{2}(\lambda_{j})\) in the basis of \(g_{*}(\lambda_{j})\) as 
\begin{equation}
  g_{1}(\lambda_{j})=\left[
  \begin{array}{cc}
  \mu & 0 \\
   0 & \frac{1}{\mu}
  \end{array}\right], \quad
  g_{2}(\lambda_{j})=\left[
  \begin{array}{cc}
   a & b \\
   c & d 
  \end{array}\right]\mbox{  }(ad-bc=1),
\end{equation}
  the following relation
\begin{equation}
\label{eq:kankei1}
  g_{*}(\lambda_{j})=g_{1}(\lambda_{j})g_{2}(\lambda_{j})
  g_{1}^{-1}(\lambda_{j})g_{2}^{-1}(\lambda_{j})=\left[
  \begin{array}{cc}
  ad-\mu^{2}bc & ab(\mu^{2}-1) \\
  cd(\frac{1}{\mu^{2}}-1) & ad -\frac{bc}{\mu^{2}}
  \end{array}\right]
\end{equation}
  must be satisfied. Since \(g_{*}(\lambda_{j})\) is assumed to have a 
  diagonal representation as 
  \(g_{*}(\lambda_{j})=\mbox{diag}\left[
  \sigma_{j},\sigma_{j}^{-1}\right]\) and from the relation 
(\ref{eq:kankei1}),  
  we obtain 
\beq 
  a=0,\quad d=0,\quad bc=-1,
\eeq
  when  \(g_{*}(\lambda_{j})\ne 
  \mbox{\boldmath$id$}\);  \(g_{2}(\lambda_{j})\)
must have the property of (b).

 Therefore, in  the basis of \(g_{*}(\lambda_{j})\),
 we have 
\begin{equation}
\label{eq:basedia}
  g_{1}(\lambda_{j})=\left[
  \begin{array}{cc}
  \mu & 0 \\
  0 & \frac{1}{\mu}
  \end{array}\right] \mbox{  and} \quad
  g_{2}(\lambda_{j})=\left[
  \begin{array}{cc}
  0 & \beta \\
  -\frac{1}{\beta} & 0
   \end{array}\right].
\end{equation}
These relations (\ref{eq:basedia}) result in  
\begin{equation}
\label{eq:comrel}
 \left[
 \begin{array}{cc}
 \sigma_{j} & 0 \\
 0 & \sigma^{-1}_{j}
 \end{array}\right]=
 g_{*}(\lambda_{j})=g_{1}(\lambda_{j})
g_{2}(\lambda_{j})g_{1}^{-1}(\lambda_{j})
 g_{2}^{-1}(\lambda_{j})=g_{1}^{2}(\lambda_{j}),
\end{equation}
where 
\(\sigma_{j}=
-\mbox{exp}\{\pi i\sqrt{25-
24\mbox{cos}\frac{j\pi}{n+1}}\}\).
However, the relation (\ref{eq:comrel}) 
causes again a contradiction  with the fact that 
the eigenvalues of \(g^{2}_{1}(\lambda_{j})\) approach to 
\beq
\{\mbox{exp}(i\pi(2\sqrt{2}\mbox{sin}(\frac{j\pi}{2(n+1)})),
\mbox{exp}(-i\pi(2\sqrt{2}\mbox{sin}(\frac{j\pi}{2(n+1)}))
\}
\eeq 
in the limit \(\epsilon\rightarrow 0\) and   there is a difference between 
the eigenvalues of \(g_{*}(\lambda_{j})\) and the eigenvalues of 
\(g^{2}_{1}(\lambda_{j})\) as will be  shown in 
the Appendix.  
 
We have seen that \(g_{1}(\lambda_{j})\) for any 
\(j(\ne \frac{n+1}{2}) \) has neither
the property of (a) or the property of (b).
Now the theorem holds.
(\bsquare {\bf End of proof of Theorem \ref{th3:fputheorem1}})\\

\section{Classification  of non-homogeneous nonlinear lattices}
\setcounter{equation}{0}
 Here in the present section, we consider the classification of  
these non-integrable nonlinear lattices via the degree of non-integrability. 
We define the degree of non-integrability  as follows:
\newtheorem{df}{Definition}
\begin{df} 
We write a relation 
\beq
  \{H,\epsilon\} \sim \{H',\epsilon'\}
\eeq
 if and only if  
 Hamiltonian systems  with \(n\) degrees of freedom \(H\) and \(H'\)
 have the same number of additional analytic integrals which are functionally 
independent together with the Hamiltonians \(H\) and \(H'\) respectively. 
Here \(\epsilon(\epsilon')\) denotes  the total energy of the 
Hamiltonian systems \(H(H')\).
\end{df}
 If we associate the positive integer
\beq 
\varrho =  
2^{r}\cdot 3^{n-r-1}
\eeq
with a Hamiltonian system \(H\) 
with \(n\) degrees of 
freedom which has \(r\) additional analytic integrals which are functionally 
independent together with the Hamiltonian \(H\), we can classify 
Hamiltonian systems via the degree of non-integrability as follows:
\beq
  \{H,\epsilon\}\sim 
  \{H',\epsilon'\} \iff \varrho(\{H,\epsilon\})=\varrho(\{H',\epsilon'\}).
\eeq 
 For a non-homogeneous nonlinear lattice  
\beq
\label{eq:nnnl2m}
  H_{\mu_{2},\mu_{4},\cdots,\mu_{2m}}(\mbox{\boldmath$q,p$};t)
=\frac{1}{2}\sum_{i=1}^{n}p^{2}_{i}+\sum_{k=1}^{m}\frac{\mu_{2k}}{2k}
   \sum_{i=1}^{n}(q_{i-1}-q_{i})^{2k}
\eeq
 of degree \(2m\), 
consider  the following transformation \(\Re \) from 
  a non-homogeneous nonlinear lattice to 
  another  non-homogeneous nonlinear lattice: 
\begin{df}
 We call a transformation \(\Re \) from  
  \(H_{\mu_{2},\mu_{4},\cdots,\mu_{2m}}(\mbox{\boldmath$q,p$};t)\) to 
  \(H_{\mu'_{2},\mu'_{4},\cdots,\mu'_{2m}}(\mbox{\boldmath$q',p'$};t')\)
  a {\bf homogenizer} 
   if the relations  
\begin{equation}
\begin{array}{l}
\label{eq:homogenizer} 
  t'=
 {\alpha}t, 
  \quad\mbox{\boldmath$q'$}= \frac{1}{\alpha^{\frac{1}{m-1}}}
  \mbox{\boldmath$q$},
  \quad\mbox{\boldmath$p'$}= 
  \frac{1}{\alpha^{\frac{m}{m-1}}}\mbox{\boldmath$p$},\\
  \mu'_{2}=\frac{1}{\alpha^{2}}\mu_{2},\quad
   \mu'_{4}=\frac{1}{\alpha^{2\frac{m-2}{m-1}}} \mu_{4},\quad \cdots,\quad 
   \mu'_{2\nu}=\frac{1}{\alpha^{2\frac{m-\nu}{m-1}}}, \quad\cdots,\quad 
   \mu'_{2m}=\mu_{2m}, 
\end{array}
\end{equation} are satisfied
 where \(\alpha\) is a real quantity larger than unity \((1<\alpha <\infty)\).
\end{df}

{\it Remark 1}

The transformation \(\Re\) in Eqs. (\ref{eq:homogenizer})
 can be regarded as a generalization of 
the scaling transformation for homogeneous nonlinear lattices 
\(H_{\mu_{2}=0,\cdots,\mu_{2m-2}=0,\mu_{2m}}\). 
Moreover, the equations of motion  
itself remain invariant under the transformation \(\Re\),   
 while a Hamiltonian is changed according to 
 the variations  of coupling constants 
\(\{\mu_{2\gamma}|1 \leq \gamma <m\}\). Thus, we can say that \(\Re\)
preserves the degree of non-integrability as 
\beq
   \{ \Re H,\Re \epsilon \} \sim \{ H,\epsilon\}.
\eeq

 {\it Remark 2}

 The operation \(\Re\) corresponds to a kind of 
{\it renormalization group} along 
 the time axis from the view point 
of the coarse graining in statistical physics 
 and it is easy to see that the fixed point of \(\Re\) 
  is the homogeneous nonlinear lattice \(H^{*}\), i.e., 
\beq
   H^{*}=\Re H^{*},
\eeq    
  where 
\beq
\label{eq:fixh} 
\begin{array}{l}
\label{eq:hhhh}
   H^{*}=\lim_{l \rightarrow \infty} \Re^{l} H\\
        =
  \frac{1}{2}\sum_{i=1}^{n}p^{2}_{i}+\frac{\mu_{2m}}{2m}
   \sum_{i=1}^{n}(q_{i-1}-q_{i})^{2m}
\end{array} 
\eeq
 and \(\Re^{l}\) denotes  \(l\) iterations of the 
operation \(\Re\).
This is the reason why the \(\Re\) is called a {\it homogenizer} here.\\

{\it Remark 3}

Under the transformation \(\Re\), the total energy \(\epsilon\) is changed 
into \(\epsilon'\) by the formula 
\beq
 \epsilon'= \Re \epsilon =\frac{1}{\alpha^{\frac{2m}{m-1}}}\epsilon.
\eeq

By the above remarks, 
we can conclude that
\beq
\label{eq:nhh}
  \{H_{\frac{1}{\alpha^{2l}}\mu_{2},\mu_{4}},\frac{1}{\alpha^{4l}}\epsilon\}
  \sim 
  \{H_{\mu_{2},\mu_{4}},\epsilon\}
\eeq
for an arbitrary FPU lattice \(\{H_{\mu_{2},\mu_{4}},\epsilon\}\).
If we take the limit 
 \(\alpha^{l}\rightarrow \infty\) 
 preserving the relation \(\frac{\epsilon}{\alpha^{4l}}\approx O(1)\),   
 L.H.S. of (\ref{eq:nhh}) approaches a homogeneous 
nonlinear lattice of degree 4, which means that 
the  degree of non-integrability of a homogeneous nonlinear lattice is
 the 
same as the degree of non-integrability of a FPU lattice in the high energy 
limit \(\epsilon\rightarrow\infty\). In relation to the non-integrability 
of the homogeneous nonlinear lattice, there is a known result 
by Yoshida\cite{yo} stating 
as follows:\\

Considering \(n\) number
of distinct exponents
\beq
\label{eq:3sin2}
\Delta_{j}\equiv
\sqrt{1+\frac{8m(2m-1)}{(m-1)^{2}}\mbox{sin}^{2}(\frac{j\pi}{2(n+1)})}
 \quad(\mbox{for}\quad j=1,\cdots,n),
\eeq
 or the \(n\) representative of the {\it Kowalevski exponents}\cite{3yo1} of 
 the homogeneous nonlinear lattice of degree \(2m\).
If the n quantities \(\{\Delta_{j}\}\) in (\ref{eq:3sin2}) 
are rationally independent,
then the  
homogeneous nonlinear lattices  have no  analytic first 
integrals except the Hamiltonian itself for \(2m(\geq 4)\) and
for odd \(n(\geq 3)\).\\

Note  that the \(n\) quantities (\ref{eq:3sin2}) equal the phase factors 
(\ref{eq:generalmonodromy}) of the eigenvalues of the commutator \(g_{*}\). 
Thus, the non-resonance hypothesis 
proving the non-integrability in the  
 low energy limit \(\epsilon \rightarrow 0\) 
induces also the non-integrability 
of nonlinear lattices with a nonhomogeneous potential function 
in the high energy limit \(\epsilon \rightarrow \infty\). 
The reason for this uniqueness of the non-resonance hypothesis 
lies in the fact that 
after a conformal mapping \(z=\phi^{4}(t)\), 
the closed loop of the commutator \(g_{1}g_{2}g^{-1}_{1}
g^{-1}_{2}\) becomes a homotopy  equivalent loop which corresponds 
to Yoshida's construction of the monodromy matrix \(G=(G_{0}G_{1})^{16}\) 
giving the same non-resonance hypothesis in terms of Kowalevski exponents.
Here, \(G_{0}\) corresponds to 
a counterclockwise 
closed loop around the singularity \(z=0\) and \(G_{1}\) corresponds to 
another counterclockwise  closed loop around the singularity \(z=1\) in the complex \(z\) plane, where 
\(z\) satisfies the following Gauss hyper geometric equation:
\beq
  z(1-z)\frac{d^{2}\xi'_{j}}{dz^{2}}+\left[\frac{3}{4}-\frac{5}{4}z\right]
 \frac{d\xi'}{dz}+\frac{3}{2}\gamma_{4}\mbox{sin}^{2}(\frac{j\pi}{2(n+1)})
 \xi'_{j}=0
\eeq
into 
which the conformal mapping \(z=\phi^{4}\) transforms the variational 
 Eq. (\ref{eq:cve}) with 
the condition \(\gamma_{2}=0\) and \(G_{0},G_{1}\) are the two 
fundamental monodromy groups of the Gauss hyper geometric equation.  
See also Fig. 3.\\ 

So far, we have succeeded in checking the non-integrability of the FPU 
lattices both in the low energy limit and in the high energy limit from a 
single non-resonance hypothesis. However, there still remain some  
problems about the integrability of FPU lattices for a finite 
energy \(0<\epsilon <\infty\). To consider the problem, we introduce a 
real parameter  
\beq
  \chi=\frac{\epsilon \mu_{4}}{(n+1)\mu^{2}_{2}}>0,
\eeq
where \(\chi\) is found to be a dimensionless parameter, because 
\beq
  \mbox{Dimension of }\chi=
  \frac{\left[\frac{\mbox{L}^{2}}{\mbox{MT}^{2}}\right]
  \left[\frac{1}{\mbox{MT}^{2}\mbox{L}^{2}}\right]}
{\left[1\right]\left[\frac{1}
  {\mbox{M}^{2}\mbox{T}^{4}}\right]}=\left[1\right] 
\eeq  
by the standard unit system as 
\(\left[\mbox{M}\right]=\left[\mbox{kg}\right],
\left[\mbox{T}\right]=\left[\mbox{sec}\right],
\left[\mbox{L}\right]=\left[\mbox{m}\right]\).
 We remark here that \(\chi\) is invariant under the transformation 
 \(\Re\) preserving the degree of non-integrability. This means that 
it is sufficient for us to classify the degree of non-integrability 
of the FPU lattices by using  
this  dimensionless parameter \(\chi\).      
Furthermore, with the use of  the parameter \(\chi\), 
we can replace the formula of the modulus 
of the elliptic function \(k\)  
 in Eq. (\ref{eq:modulus}) by 
\beq
k=\frac{1}{\sqrt{2}}
\sqrt{1-\frac{1}{\sqrt{1+4\chi}}}.
\eeq
 We note here that \(k\rightarrow 0\) if and only if 
 \(\chi \rightarrow 0\) and this 
 limit can be realized for each set of \(\{\epsilon,\mu_{2},\mu_{4}|
 0<\epsilon,\mu_{2},\mu_{4}<\infty\}\) when we take \(n\)(:odd) a fully 
great number. If we recall that in Theorem 1  
it is essential 
for proving the
non-integrability of the FPU lattices
to take the limit \(k\rightarrow 0(\chi\rightarrow 0)\) 
so that 
the eigenvalues of \(g_{1}(\lambda_{j})\) approach the quantities 
in Eq. (\ref{eq:limiteigenvalues}), 
 we have the following theorem:    

\begin{th3}
\label{th3:fpufinal} 
A  
FPU lattice(\ref{eq:ulam}) characterized by 
 an arbitrary set of the system parameters 
\(\{\epsilon,\mu_{2},\mu_{4}|0<\epsilon<\infty, 0<\mu_{2}<\infty,
0<\mu_{4}<\infty\}\) 
 has no additional  analytic 
integrals of motion besides the Hamiltonian itself for
a fully great  number of degrees of freedom \(n(n:\mbox{odd})\). 
\end{th3}

This result shows that we need 
a fully great number \(n\) of degrees of freedom to guarantee the strongest 
non-integrability  \(\varrho(\{H_{FPU},\epsilon\})=3^{n-1}\)
 of each FPU lattice \(H_{FPU}\) with the systems parameters 
\(\{\mu_{2},\mu_{4}|0<\mu_{2}<\infty,0<\mu_{4}<\infty\}\) for    
\(0<\epsilon<\infty\) and a problem remains still open as to 
 whether we can prove
 the non-integrability 
 of each FPU lattices with the system parameters 
\(\{\mu_{2},\mu_{4},n|0<\mu_{2}<\infty,0<\mu_{4}<\infty,
n:\mbox{odd},n\geq 3\}\) 
for a given finite energy \(\epsilon(0<\epsilon<\infty)\).   
Clearly, Theorem 1 and Theorem 2 have the reciprocal conditions 
in terms of 
the system parameters \(\{\epsilon,\mu_{2},\mu_{4},n\}\) 
and we can 
unify them into a single theorem as follows: 
\begin{th3}
A FPU lattice(\ref{eq:ulam}) characterized by the system parameters\\ 
\(\{\epsilon,\mu_{2},\mu_{4},n|0<\epsilon<\infty,0<\mu_{2}<\infty,
0<\mu_{4}<\infty,
n\geq 3, n:\mbox{odd}
\}\) has no additional analytic integrals of motion besides the 
Hamiltonian itself when the following condition 
\beq
\chi=\frac{\epsilon \mu_{4}}{(n+1)\mu^{2}_{2}}\rightarrow 0 
\eeq   
is satisfied.

\end{th3}
\section{Summary and discussion}
We have got the non-integrability proof 
 of the FPU lattices  
 by tracing the following steps. 
In Section 2, we have shown that the singularity analysis can be performed towards 
 the non-homogeneous nonlinear lattices by finding the special 
straight-line solutions in terms of elliptic functions.  
Especially, it is shown there that the normal variational equations 
can be decoupled into  separated equations   
and each of them happens to be   
a  Lam\'e equation and consequently  we can compute 
the eigenvalues of the monodromy 
matrices associated with the counterclockwise loop along the 
period cell of the special solutions. 
  In  Section 3, with the use of Yoshida's theorem whose  validity is 
guaranteed when  
that the normal variational equations can be decoupled, 
 we can prove the non-existence of the additional analytic conserved 
quantities 
besides the Hamiltonian itself 
for  FPU lattices in the low energy limit.
In Section 4,  
we have considered the classification of FPU lattices via the  
  degree of non-integrability. There, by introducing  a transformation 
\(\Re\) from a non-homogeneous nonlinear lattice 
into another  non-homogeneous nonlinear lattice, 
which preserves the equations of motion, we have shown that the degree of 
non-integrability of each 
FPU lattice in the high energy limit is the same as that 
 of the homogeneous nonlinear lattice whose non-integrability
is known to be proven by the analysis   
using  Kowalevski exponents.
In case of FPU lattices with  arbitrary energy in \(0<\epsilon<\infty\), 
 we have a theorem showing  the non-integrability for a fully great 
number of degrees of freedom 
by introducing a dimensionless parameter which characterizes the 
normal variational equations. This theorem (Theorem 2) and Theorem 1 
have the reciprocal sufficient conditions for the non-integrability of 
FPU lattices and finally they are  unified into a single theorem (Theorem 3) 
 by the  dimensionless parameter \(\chi\).

There remains an open problem to be solved:
Can we have a theorem about the non-integrability of more general 
nonlinear lattices of degree \(2m>4\)?  More specifically speaking, can   
  we have a theorem asserting the non-integrability of the 
non-homogeneous nonlinear lattices of degree \(2m>4\) such as Theorem 1 and 
Theorem 2? 
The present neck is in a fact that   
the normal variational equations along the special solutions 
of {\it hyper-elliptic functions} 
do not belong to a class of Lam\'e equations,
though they can always be decoupled into \(n-1\) 
separated variational equations,
and 
we cannot compute the eigenvalues of the 
monodromy matrices associated with these variational equations 
in general  and therefore we cannot perform
 the Ziglin analysis completely. 
 However, the  problem on the degree of 
non-integrability for  more general
non-homogeneous nonlinear lattices remains an interesting 
open question to be solved,  because  
the present analysis on the non-integrability of the FPU lattices, 
together with Yoshida's argument about the homogeneous 
nonlinear lattice in terms of  Kowalevski exponents, strongly suggests 
 universality about the non-integrability 
of more general non-homogeneous nonlinear lattices in Eq. 
(\ref{eq:nnnl2m}).\\

\section{Appendix}
\setcounter{equation}{0}
In this appendix, we show the difference between the eigenvalues of 
  \(g_{*}(\lambda_{j})\) and the eigenvalues of \(g^{2}_{1}(\lambda_{j})\)
for any \(j(\ne\frac{n+1}{2})\). 
We prove 
this by reductio ad absurdum. First, we  assume that the following equality  
\beq
\label{eq:heq}
\begin{array}{l}
  \mbox{Spec} (g^{2}_{1}(\lambda_{j}))\equiv
\{\mbox{exp}(i\pi(2\sqrt{2}\mbox{sin}(\frac{j\pi}{2(n+1)})),
\mbox{exp}(-i\pi(2\sqrt{2}\mbox{sin}(\frac{j\pi}{2(n+1)}))
\}\\
= 
\{-\mbox{exp}(\pi i\sqrt{25-
24\mbox{cos}\frac{j\pi}{n+1}}),
-\mbox{exp}(-\pi i\sqrt{25-
24\mbox{cos}\frac{j\pi}{n+1}})\}
\equiv \mbox{Spec}(g_{*}(\lambda_{j}))
 
\end{array}
\eeq 
would hold for some \(j(\ne\frac{n+1}{2})\). 
Our purpose here is to show that the relation (\ref{eq:heq}) is false. 
By (\ref{eq:heq}),  
 one of the following equalities 
\beq
\label{eq:eqp}
\sqrt{25-
24\mbox{cos}\frac{j\pi}{n+1}}+1+2m_{+}=
2\sqrt{2}\mbox{sin}(\frac{j\pi}{2(n+1)})
\eeq
\beq
\label{eq:eqm} 
\sqrt{25-
24\mbox{cos}\frac{j\pi}{n+1}}+1+2m_{-}=
-2\sqrt{2}\mbox{sin}(\frac{j\pi}{2(n+1)})
\eeq
 holds for \(m_{+},m_{-}\in \mbox{\boldmath$Z$}\) and 
  \(j\ne\frac{n+1}{2}\). 
Because of the inequalities 
\beq
1<\sqrt{25-
24\mbox{cos}\frac{j\pi}{n+1}}<7,\quad 
0<2\sqrt{2}\mbox{sin}(\frac{j\pi}{2(n+1)})<2\sqrt{2},
\eeq
it is sufficient to consider the equalities (\ref{eq:eqp}) and 
(\ref{eq:eqm}) for 
\(m_{+}\in \{-3,-2,-1,0\}\) and \(m_{-}\in \{-5,-4,-3,-2\}\).
If we rewrite \(\mbox{sin}(\frac{j\pi}{2(n+1)})\) as \(X\), 
 Eq. 
(\ref{eq:eqp}) and Eq. (\ref{eq:eqm}) 
are changed into the following equations 
\beq
40X_{+}^{2}+(1+2m_{+})4\sqrt{2}X_{+}+1-(1+2m_{+})^{2}=0,
\eeq
\beq
40X_{-}^{2}-(1+2m_{-})4\sqrt{2}X_{-}+1-(1+2m_{-})^{2}=0
\eeq
respectively. 
By considering \(X(=X_{+},X_{-}\geq 0\), 
we can easily obtain a series of  the solutions as 
follows:
\beq
\begin{array}{l}
X_{+}(m_{+}=0)=0\\ 
X_{+}(m_{+}=-1)=\frac{\sqrt{2}}{10}\\
X_{+}(m_{+}=-2)=\frac{1}{\sqrt{2}}\\
X_{+}(m_{+}=-3)=\frac{5\sqrt{2}+\sqrt{290}}{20}>\frac{7+\sqrt{289}}{20}=
  \frac{6}{5}>1\\
X_{-}(m_{-}=-2)=\frac{\sqrt{2}}{5}\\
X_{-}(m_{-}=-3)=\frac{-5\sqrt{2}+\sqrt{290}}{10}\\
X_{-}(m_{-}=-4)=\frac{1}{\sqrt{2}}\\
X_{-}(m_{-}=-5)=\frac{-9\sqrt{2}+\sqrt{962}}{20}.
\end{array}
\eeq
Furthermore, because of the following relations  
\(0<X_{\pm}<1,\quad X_{\pm}\ne \frac{1}{\sqrt{2}}\) for 
\(j \ne \frac{n+1}{2}\),  
we can discard the solutions \(X_{+}(m_{+}=0),X_{+}(m_{+}=-2),X_{+}(m_{+}=-3),
X_{-}(m_{-}=-4)\). For the solutions 
\(X_{+}(m_{+}=-1),X_{-}(m_{-}=-2)\), we have the equalities as
\beq
\label{eq:relap1}
\begin{array}{l}
\mbox{cos}\frac{j\pi}{n+1}=1-2(X_{+}(m_{+}=-1))^{2}=\frac{24}{25}\\
\mbox{cos}\frac{j\pi}{n+1}=1-2(X_{-}(m_{-}=-2))^{2}=\frac{21}{25}
\end{array}
\eeq
respectively. However, 
these relations (\ref{eq:relap1}) contradict with a fact
\cite{4um0}  that   
\beq
\mbox{cos}\frac{j\pi}{n+1}\in \mbox{\boldmath$Q$} \iff
  \mbox{cos}\frac{j\pi}{n+1}=0,\quad
\mbox{or}\quad \frac{1}{2},\quad \mbox{or}\quad -\frac{1}{2}.
\eeq
Thus, we can also discard the solutions \(X_{+}(m_{+}=-1)\) and 
\(X_{-}(m_{-}=-2)\).
Let us consider the solutions \(X_{-}(m_{-}=-3)\) and \(X_{-}(m_{-}=-5)\).
If we rewrite \(\mbox{exp}\frac{j\pi}{n+1}\) as \(\zeta\),
the following relation    
\beq
\label{eq:relap2}
  25\zeta^{4}+70\zeta^{3}-46\zeta^{2}+70\zeta+25=0
\eeq
is  satisfied for the solution \(X_{-}(m_{-}=-3)\),
while  the relation 
\beq
\label{eq:relap3}
  25\zeta^{4}+462\zeta^{3}+626\zeta^{2}+462\zeta+25=0
\eeq
is satisfied for the solution \(X_{-}(m_{-}=-5)\).
 Since \(\zeta\) is one of \(2(n+1)\)-th root of unity and 
 \(2(n+1)\quad (\mbox{mod} \quad 4)=0\), the minimal polynomial \(P(\zeta)\) 
with  coefficients of 
rational integers which has a solution \(\zeta\) for \(P(\zeta)=0\) 
must be one of the {\it cyclotomic polynomials} \(\Phi_{l}(Y)\in 
\mbox{\boldmath$Z$}[Y]\):
\beq
  \Phi_{l}(Y)=\prod_{i=1}^{\varphi(l)}(Y-\zeta_{i}),
\eeq
 where \(2(n+1) (\mbox{ mod }l) =0\), 
\(\{\zeta_{i}\}\) is  a set of all primitive \(l\) th roots of 
unity and \(\varphi(l)\) is the number of positive integers which are less 
than or equal to \(l\) and relatively prime to \(l\) and this 
function \(\varphi\) is called 
Euler's \(\varphi\)-function.    
All cyclotomic polynomials \(\Phi_{l}(Y)\) are known to be 
irreducible over \(\mbox{\boldmath$Q$}\) and we can check that 
 cyclotomic polynomials 
\(\Phi_{l}(Y)\) whose degrees are less than or equal to \(4\)
 are restricted 
to the case\cite{4umth} that 
\beq 
\begin{array}{l} 
 \aleph \equiv \{ \Phi_{l}(Y)|2\leq\mbox{Deg}\Phi_{l}(Y)\leq 4\}\\ = 
\{\Phi_{3}(Y)=Y^{2}+Y+1,\Phi_{4}(Y)=Y^{2}+1,
 \Phi_{5}(Y)=Y^{4}+Y^{3}+Y^{2}+Y+1,\\
\Phi_{6}(Y)= Y^{2}-Y+1,
 \Phi_{8}(Y)=Y^{4}+1,
 \Phi_{12}(Y)=Y^{4}-Y^{2}+1\}.
\end{array}
\eeq  
Using this fact, we can say  that  a cyclotomic polynomial   
\(\Phi_{l}(Y)\in \aleph\) must divide either 
the  polynomial
\beq
  P_{1}(Y)\equiv 25Y^{4}+70Y^{3}-46Y^{2}+70Y+25
\eeq
in Eq.(\ref{eq:relap2})
 or the  polynomial  
\beq
  P_{2}(Y)\equiv 25Y^{4}+462Y^{3}+626Y^{2}+462Y+25
\eeq
in 
Eq.(\ref{eq:relap3}). It is easy to check that 
this is impossible.  Thus now, we can 
conclude that the assumption (\ref{eq:heq}) is false; i.e., we have   
shown that 
for any \(j(\ne \frac{n+1}{2})\)
\beq
  \mbox{Spec}(g_{1}^{2}(\lambda_{j}))\ne \mbox{Spec}(g_{*}(\lambda_{j})).
\eeq\\

\Large 
{\bf Acknowledgements}
\normalsize\\
 
The majority of this work was carried out to complete 
the doctoral dissertation at Department of Physics,
University of Tokyo. 
 I would 
like to thank 
Prof. Masuo Suzuki, Prof. Miki Wadati and Prof. Haruo Yoshida for 
 valuable discussions. 
 While writing this work up 
I have  been supported   
   from  
 the Special Researcher's Program towards  Basic 
  Science at the RIKEN
 and from the Program 
 of the Complex Systems II at the International Institute for 
 Advanced Study (IIAS-Kyoto). I would like to thank Prof. Shun-ichi 
Amari for his  
continual encouragement.

\end{document}